\documentclass[a4paper,11pt]{article}
\usepackage{pos}
\usepackage{graphicx}
\usepackage{subcaption}

\title{Study of $\mathrm{^{26}Al}$ in the COSI 2016 superpressure balloon flight}
 \ShortTitle{$\mathit{^{26}Al}$ in the COSI 2016 SPB flight}

\author*[a]{Jacqueline Beechert}

\affiliation[a]{Space Sciences Laboratory, University of California, Berkeley,\\
  7 Gauss Way, Berkeley, CA 94720, USA}

\forColl{COSI} 

\emailAdd{jbeechert@berkeley.edu}

\abstract{The Compton Spectrometer and Imager (COSI) is a balloon-borne compact Compton telescope designed to survey the $\gamma$-ray sky from 0.2 to 5 MeV. COSI's wide field-of-view (FOV) and excellent energy resolution from high-purity germanium detectors make it uniquely capable of probing this under-explored energy regime. In particular, it can facilitate understanding of stellar nucleosynthesis through studies of diffuse emission from the radioisotope $\mathrm{^{26}Al}$ at 1.809 MeV. In 2016, COSI was launched from Wanaka, New Zealand on a NASA superpressure balloon and flew for 46 days. The flight was a technologic and scientific success, boasting live detection and polarization studies of GRB160530A, spectral analysis of the Crab Nebula and the 511-keV positron annihilation emission at the Galactic Center, and detection of Cygnus X-1. This article details the first maximum-likelihood search for the 1.809 MeV signature of Galactic $\mathrm{^{26}Al}$ in the 2016 data. The analysis reveals a promising excess around the expected energies of an $\mathrm{^{26}Al}$ signature with 3.7$\sigma$ significance and a measured flux of (17.0 $\pm$ 4.9) $\times$ 10$^{-4}$ ph cm$^{-2}$ s$^{-1}$. Further exploration is currently underway to solidify the measurement.}

\FullConference{37$^{\rm{th}}$ International Cosmic Ray Conference (ICRC 2021)\\
		July 12th -- 23rd, 2021\\
		Online -- Berlin, Germany}

\begin{document}
\maketitle

\section{Introduction} \label{sec:introduction}

Aluminum-26 ($\mathrm{^{26}Al}$) is a radioactive isotope that decays to an excited state of Magnesium-26 ($\mathrm{^{26}Mg}$) with a half-life time of 0.7 Myr. The excited $\mathrm{^{26}Mg}$ de-excites to its ground state, emitting a 1.809 MeV $\gamma$-ray. Because of its long-lived nature, $\mathrm{^{26}Al}$ is likely to decay after ejection from its stellar production sites into the interstellar medium (ISM), facilitating both studies of the stellar conditions responsible for nucleosynthesis and the hot phase of the ISM. Hence, $\mathrm{^{26}Al}$ traces the life-cycle of elements in the Milky Way from their initial synthesis to their feedback into the ISM. 

The 1.809 MeV line was first detected in the Milky Way by the High Energy Astronomy Observatory (HEAO-3) satellite in 1984 \citep{mahoney1984heao}. The Compton Telescope (COMPTEL) on board the Compton Gamma-Ray Observatory (CGRO) advanced studies of $\mathrm{^{26}Al}$ in the 1990s by mapping the diffuse emission \citep{pluschke2001comptel}, and the spectrometer SPI on board the International Gamma-ray Astrophysics Laboratory (INTEGRAL) satellite is currently measuring the signal. 

SPI has mapped $\mathrm{^{26}Al}$ \citep{Bouchet_2015} and detected the line at 1809.83 $\pm$ 0.04 keV. The full-sky flux was found to be (1.7 $\pm$ 0.1) $\times$ 10$^{-3}$ ph cm$^{-2}$ s$^{-1}$ \citep{siegert2017positron}. Kinematic measurements reveal a systematic shift of the line with Galactic longitude, implying a velocity of $\mathrm{^{26}Al}$ emission up to 200 km s$^{-1}$ greater than that expected from Galactic rotation \citep{Kretschmer_2013,Dame_2001}. Ejection of stellar material carrying $\mathrm{^{26}Al}$ into merging, low-density HI supershells leading the spiral arms of the Galaxy has been proposed as an explanation of the observed velocity \citep{krause201526al}. A complete review of current understanding of $\mathrm{^{26}Al}$, with an emphasis on the contributions from SPI, is provided in \cite{diehl2020steady}. 

The Compton Spectrometer and Imager (COSI) is a balloon-borne compact Compton telescope with excellent spectral resolution of 0.24\% FWHM at 1.8 MeV, making it uniquely capable of new measurements required to resolve the primary sources of $\mathrm{^{26}Al}$ emission and its distribution throughout the ISM. In this proceedings, we search for a signal of $\mathrm{^{26}Al}$ in the COSI 2016 flight data and compare the results to expectations from previous studies of $\mathrm{^{26}Al}$. Section \ref{sec:cosi} provides an overview of the COSI 2016 flight, Section \ref{sec:search} details the analysis, and Section \ref{sec:discussion_flux} discusses the results. This study establishes the scientific potency of modern Compton telescopes in nucleosynthesis studies and is key proof-of-concept for such studies on the Compton telescope spacecraft mission, COSI-SMEX, currently in competitive NASA Phase A development \citep{tomsick2019compton}. 

\section{The Compton Spectrometer and Imager} 
\label{sec:cosi}
COSI was launched as a science payload on a NASA superpressure balloon from Wanaka, New Zealand on May 17, 2016. Twelve high-purity cross-strip Ge detectors (each 8 $\times$ 8 $\times$ 1.5 cm$^3$) arranged in a 2 $\times$ 2 $\times$ 3 array comprise a 972 cm$^3$ active volume within which incident photons of energies 0.2--5 MeV are measured. The energy and three-dimensional position of each interaction are recorded, enabling reconstruction of the photon path and a localization of the incident photon direction to a circle on the sky. Detailed explanations of calibrations, event reconstruction, and imaging algorithms employed in Compton telescopes are provided in \cite{sleatorDEE} and \cite{zoglauer2021cosi}. COSI has a wide FOV of $\sim$1$\pi$ sr defined by the five anti-coincidence CsI shields that surround the four sides and bottom of the detector array. It is always pointed at zenith and sweeps the sky through the Earth's rotation during flight. The shields actively veto the dominant $\gamma$-ray and cosmic-ray background incident from below the instrument, namely that of atmospheric Earth albedo radiation. 

COSI performed well with nine detectors operating continuously throughout the 46-day flight. Cold temperatures at night and balloon leaks caused altitude drops below the nominal flight altitude of 33 km, which exacerbated high background contamination from albedo radiation and increased atmospheric absorption \citep{kierans2016}. Extensive simulations are required to model the flight data and as a mostly unbiased approach, we treat COSI as a collimated light bucket, with the collimation performed by data selections through the Compton Data Space \cite{zoglauer2021cosi}.

\section{\texorpdfstring{Search for \boldmath{$\mathrm{^{26}Al}$} in the 2016 flight data}{Search for 26Al in the 2016 flight data}}
\label{sec:search}

\subsection{Data selection} \label{sec:data_selections}
\begin{figure}
    \centering
      \includegraphics[width=0.5\textwidth, trim=0.1in 0.03in 0in 0.0in, clip=true]{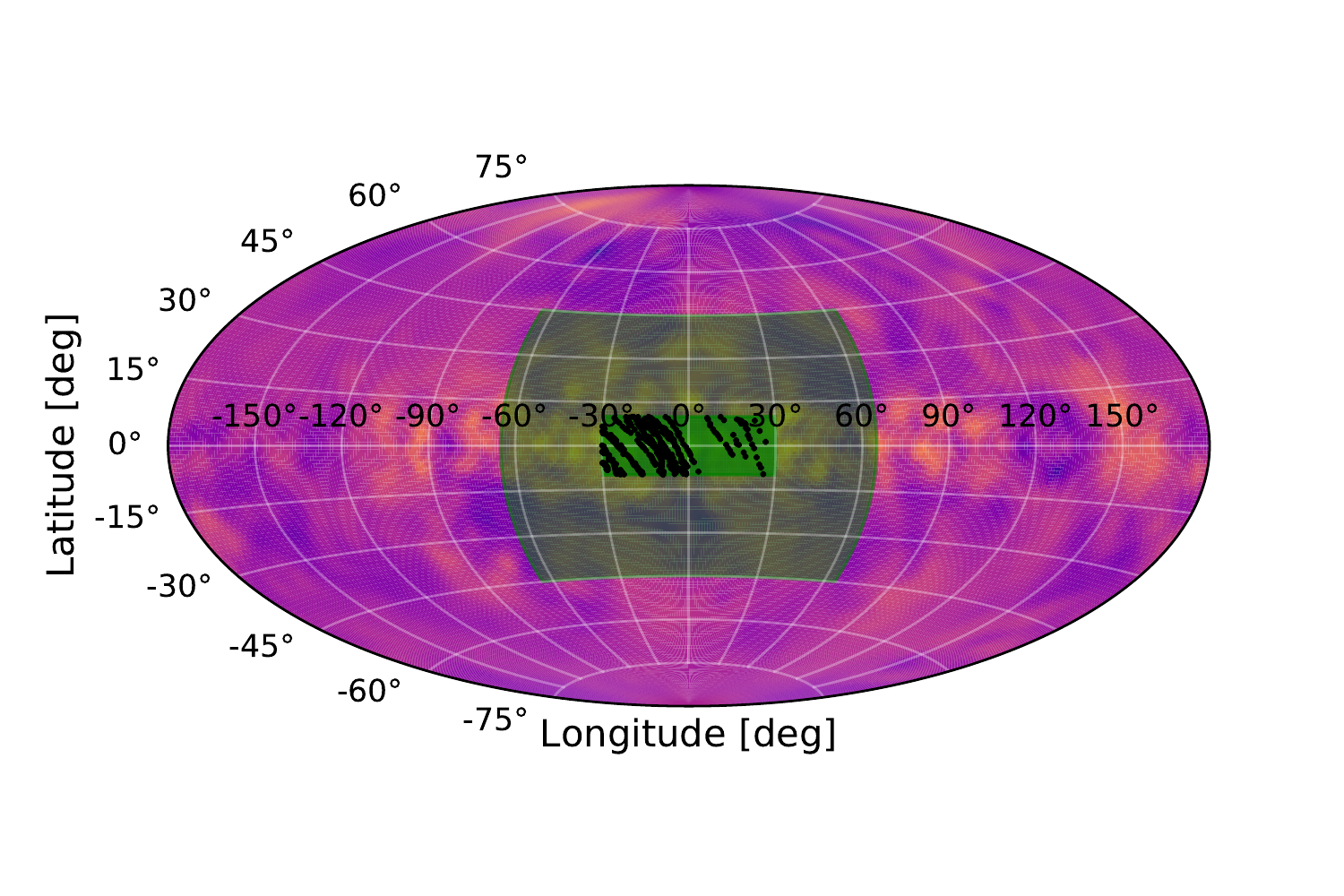}~
      \includegraphics[width=0.43\textwidth, trim=0.0in 0.0in 0.5in 0.2in, clip=true]{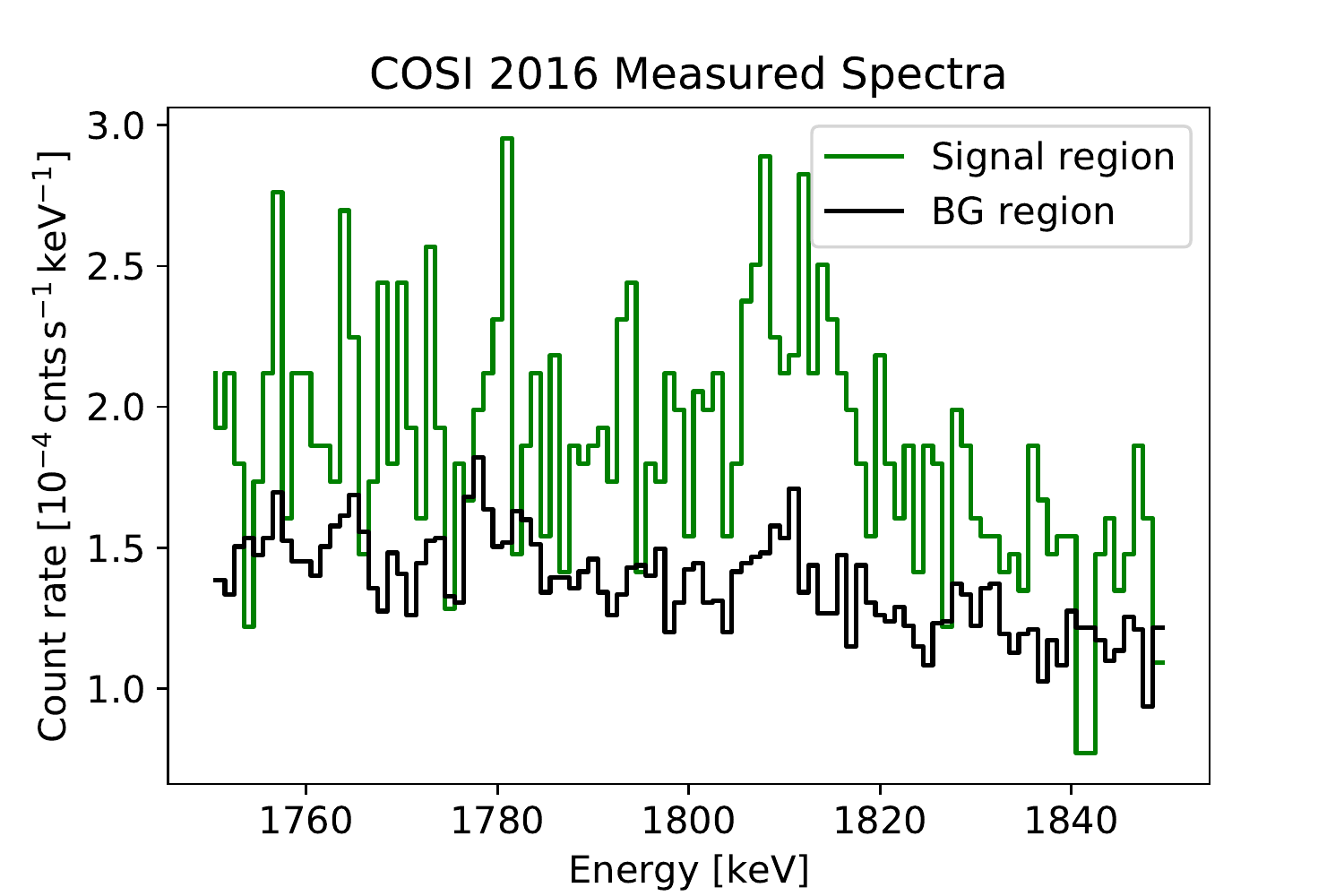}
    \caption{Left: COSI 2016 flight exposures (black points) within the Inner Galaxy pointing cuts (central green region: (|$\ell|$ $\leq$ 30$^{\circ}$, |$b|$ $\leq$ 10$^{\circ}$)) plotted over the SPI $\mathrm{^{26}Al}$ image \cite{Bouchet_2015}. The broader green region maps the maximum extension given by $\phi_{\rm max}$ = 35$^{\circ}$. Right: Flight spectra in the signal and background regions.}
    \label{fig:flight_path_and_spectra}
\end{figure}

The bulk of $\mathrm{^{26}Al}$ diffuse emission is from the Inner Galaxy and previous observations have focused on this region. Based on this, we define the signal region by a spherical rectangle, typically used for this type of analysis, with (|$\ell|$ $\leq$ 30$^{\circ}$, |$b|$ $\leq$ 10$^{\circ}$). The COSI 2016 zenith pointings in the signal region are displayed over the SPI 1.809 MeV image in Figure \ref{fig:flight_path_and_spectra}. Because of COSI's large FOV, we also expect photons outside this region to contribute to our signal spectrum. To minimize overlap between photons originating from inside and outside of the signal region, the applied pointing cuts are curated to respect the effective broadening of the regions by the incident photons' maximum allowed Compton scattering angle $\phi$. 

An optimization procedure performed on a simulated $\mathrm{^{26}Al}$ signal and simulation of atmospheric background \citep{ling_model} reveals that a minimum $\phi_{\rm min}$ = 10$^{\circ}$ removed more background than signal events. To preserve a significant fraction of the sky as background for robust background statistics, $\phi_{\rm max}$ was chosen to be 35$^{\circ}$. Since the signal is mainly concentrated in the Inner Galaxy, this selection has no consequence for the expected signal-to-noise ratio. The pointing cuts for the background region were thus chosen such that the border of each cut plus 35$^{\circ}$ fell tangential to, but not overlapping with, the 35$^{\circ}$-broadened signal region. Three such cuts were combined to form the background region: ($\ell$,$b$) = (-180$^{\circ}$ $\pm$ 80$^{\circ}$, 0$^{\circ}$ $\pm$ 90$^{\circ}$), (0$^{\circ}$ $\pm$ 30$^{\circ}$, 85$^{\circ}$ $\pm$ 5$^{\circ}$), and (0$^{\circ}$ $\pm$ 30$^{\circ}$, -85$^{\circ}$ $\pm$ 5$^{\circ}$). 

Additional event selections restricted the flight data to photons of interest: Compton events with energy 1750--1850 keV originating less than 90$^{\circ}$ from COSI's zenith (this ``Earth Horizon Cut'' reduces the dominant albedo background) were considered. Only events which scattered between two and seven times in the instrument, with minimum distance between the first two interactions of 0.5 cm and any subsequent interactions of 0.3 cm, passed selection. To mitigate the effects of increased atmospheric background and attenuation at lower altitudes, data taken in the signal region were restricted to times when the balloon floated at a minimum altitude of 33 km (see later results in Figure \ref{fig:zenith_resp_sig_vs_altitude}). Because the changing background level does not change shape with altitude, we also use data from altitudes below 33 km to determine the shape, but not the normalization, of the spectrum in the background region; the only time restrictions were the removal of events taken before the balloon achieved float altitude and when it saw high shield rates. The total observation time in the signal region was $\sim$156 ks and that in the background region was $\sim$1356 ks. Because two detectors were turned off within the first 24 hours of flight and a third on June 6, 2016, flight data from before and after June 6 were processed separately with 10- and 9-detector mass models, respectively. 

\subsection{Maximum likelihood approach}
We model the data, $d$, in our signal region as a linear combination of sky, $s$, and background, $b$, contributions with unknown amplitudes $\alpha$ and $\beta$, respectively, in 100 1-keV energy bins, $i$, spanning 1750--1850 keV. The model templates $s$ and $b$ are explained below and the total model reads:

\begin{equation}
    m_i = \alpha s_i + \beta b_i \\
    \label{eq:model}
\end{equation}

We maximize the likelihood $\mathcal{L}(d|m)$ = $\prod_{i = 1}^{N} \frac{m_{i}^{d_i}e^{-m_i}}{d_{i}!}$ for $\alpha$ and $\beta$; $d_i$ is the number of received photons in the signal region per energy bin (Figure \ref{fig:flight_path_and_spectra}).

\subsection{Sky model}
\label{sec:flightdata_sky_model}
\begin{figure}
\centering
  \includegraphics[width=0.48\linewidth]{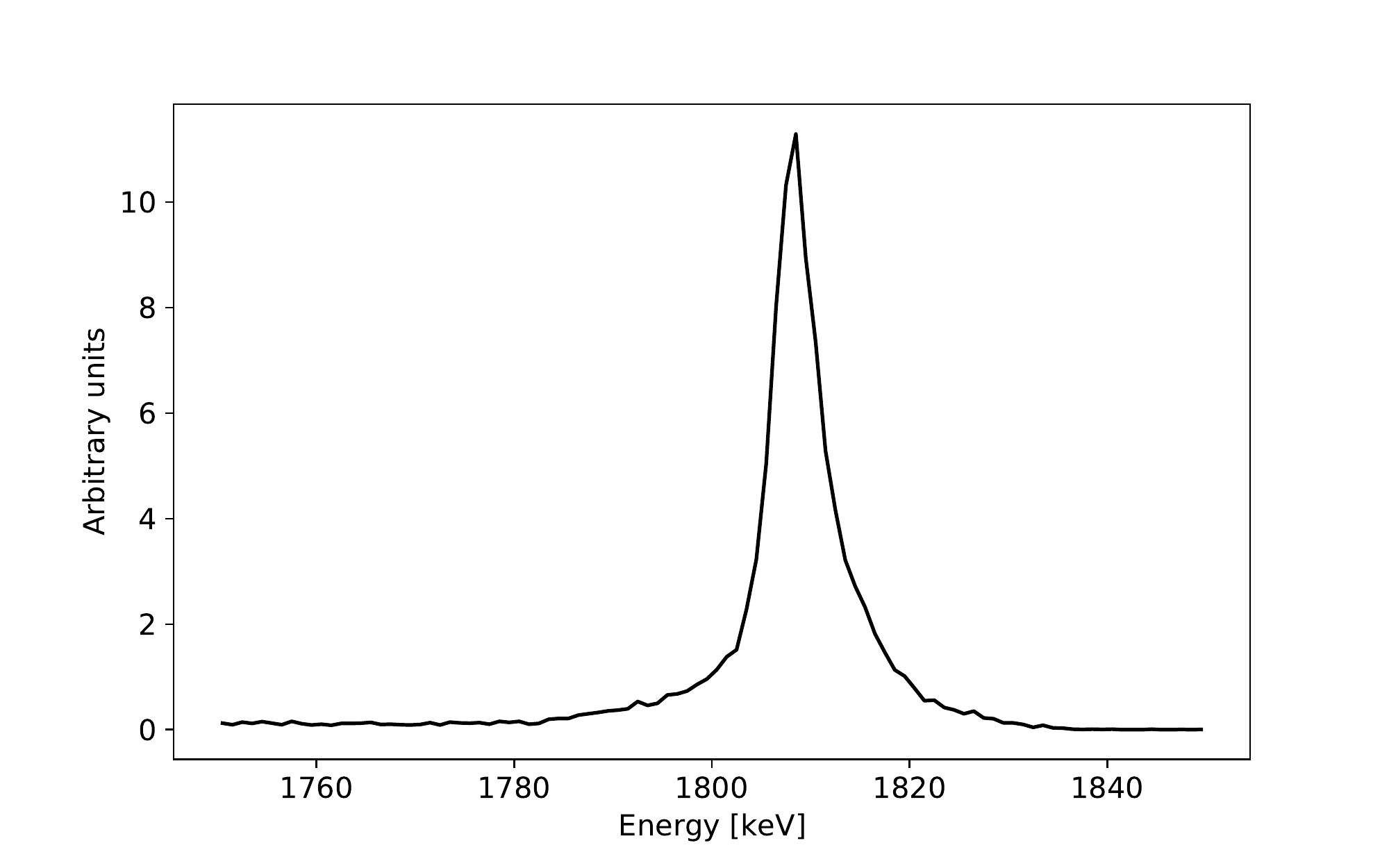}
  \includegraphics[width=0.48\linewidth]{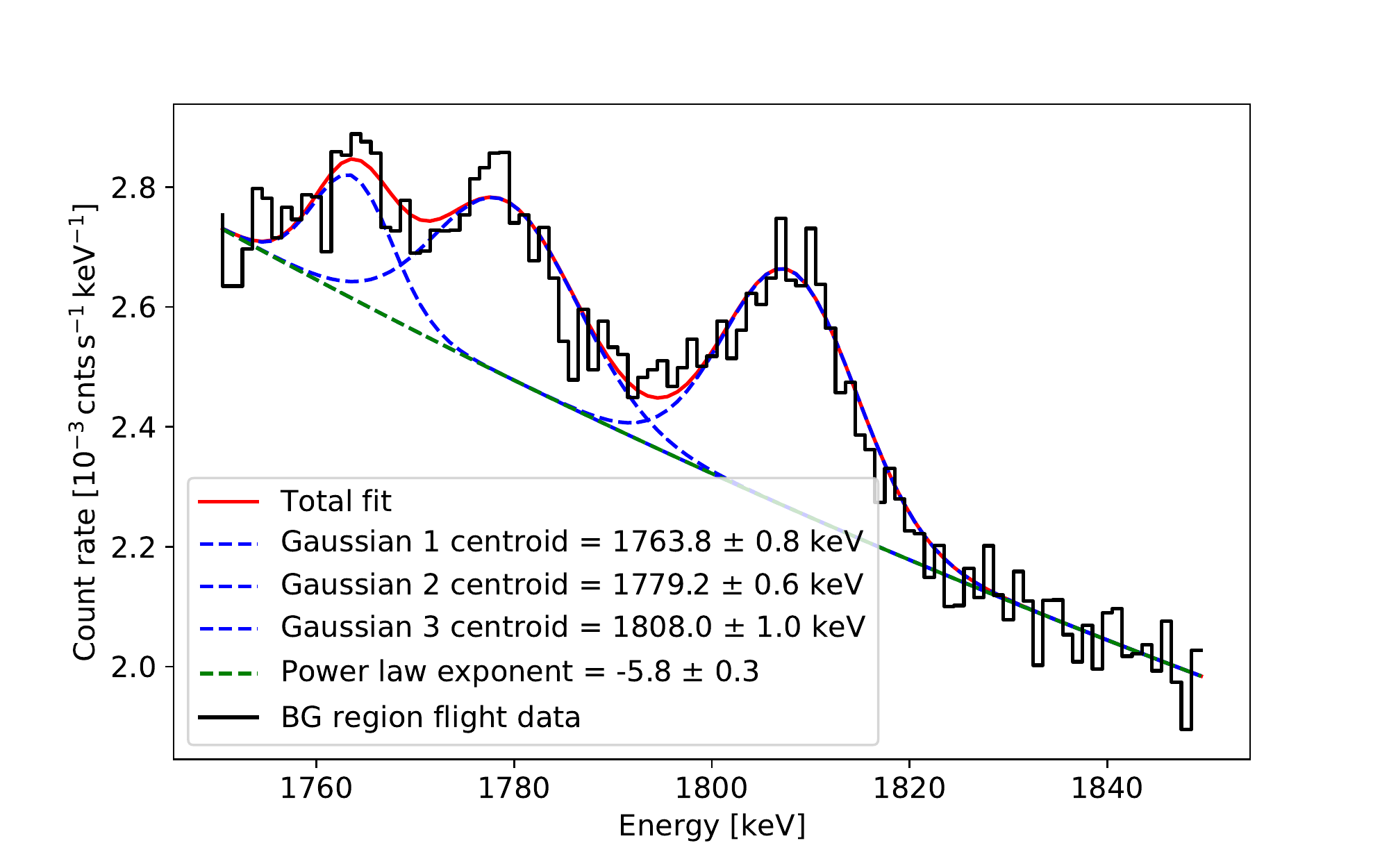}
\caption{Left: The spectral sky model is defined by fifty simulations of COSI's response to the DIRBE 240 $\mu$m map. Right: The fitted spectrum of COSI flight data in the background region with minimal event selections.}
\label{fig:sky_bg_models}
\end{figure}

The spectral response to the expected signal (e.g. SPI or COMPTEL map) results in approximately 41 photons in the signal region over the $\sim$156 ks signal region observing time. To provide a smooth response, we run 50 simulations of the COSI flight response to the DIRBE 240 $\mu$m map \citep{dirbe}. We choose the DIRBE map because the SPI and COMPTEL image reconstructions carry an intrinsic measurement bias and produce weak artifacts which cannot easily be distinguished from true emission. The DIRBE map is an adequate tracer of $\mathrm{^{26}Al}$ \citep{knodlseder1999multiwavelength,Bouchet_2015} and is more structured than the 3$^{\circ}$ resolutions of SPI and COMPTEL. The energy spectrum of these simulated events which pass the selections applied to flight data in the signal region defines the spectral sky model (Figure \ref{fig:sky_bg_models}).

\subsection{Background model}
\label{sec:flightdata_bg_model}

The background model is inferred from the data in the background region of the 2016 flight (Figure \ref{fig:flight_path_and_spectra}). Although high-latitude emission at 1.8 MeV is recently discussed in the literature \citep{pleintinger2019comparing,rodgers-lee}, COSI observations of these regions are dominated by instrumental background radiation and were thus deemed adequate for estimating the background contribution. 

To find a smoother description of the background with enhanced statistics, we generate the spectrum of the background region with minimal event selections: we consider Compton events from the background region with incident energies between 1750 and 1850 keV, Compton scattering angles ranging from 0$^{\circ}$ to 90$^{\circ}$, no minimum distance between interactions, and no Earth Horizon cut. This spectrum is fitted with a power law plus $l$ = 3 Gaussian-shaped $\gamma$-ray lines:

\begin{equation}
    b(E) = C_0 \left( \frac{E}{E_c} \right)^{\gamma} + \sum_{l = 1}^{3} \frac{A_l}{\sqrt{2\pi}\sigma_l} \exp\left(-\frac{1}{2} \left( \frac{E-E_l}{\sigma_l} \right)^2 \right)\mathrm{.}
    \label{eq:bg_model}
\end{equation}

In Equation \ref{eq:bg_model}, $C_0$ is the amplitude of the continuum, $E_c$ = 1.8 MeV is the pivotal energy, $\gamma$ is the power law index, and $A_l$, $E_l$, $\sigma_l$ are the rate, centroid, and width of Gaussian-shaped line $l$. 

Figure \ref{fig:sky_bg_models} shows the fit of Equation \ref{eq:bg_model} to the background spectrum. The fitted parameters are in Table \ref{table:flight_data_bg_priors}. The continuum is from atmospheric background and the lines $\ell$ are from instrumental background decays. We include the fitted spectral shapes and their uncertainties as priors to the signal region fit, discussed in the next section.

\begin{table}[ht]
\tabcolsep=0.22cm
\centering
\begin{tabular}{c||rr|rrr|rrr|rrr}
                      & $C_0$  & $\gamma$ & $A_1$  & $E_1$  & $\sigma_1$ & $A_2$  & $E_2$ & $\sigma_2$ & $A_3$  & $E_3$  & $\sigma_3$ \\
\hline
Value       & 23.2 & -5.8     & 2.0 & 1763.8 & 3.8        & 5.2 & 1779.2 & 7.1        & 6.6 & 1808.0 & 6.6        \\
Uncertainty  & 0.3   & 0.3      & 0.7  & 0.8    & 1.0        & 0.8 & 0.6    & 1.2        & 0.6  & 1.0    & 0.5     
\end{tabular}

\caption{Fitted parameters of the background region with minimal event selections (Figure \ref{fig:sky_bg_models}). Units: [$C_0]$ = $10^{-4}$ cnts s$^{-1}$ keV$^{-1}$, [$A_{l}$] = 10$^{-3}$ cnts s$^{-1}$, [$E_{l}$] = [$\sigma_{l}$] = keV.} 
\label{table:flight_data_bg_priors}
\end{table}

\subsection{Fitting the sky and background simultaneously}
\label{sec:flightdata_fitting_sky_and_bg}
We fit the sky and background models simultaneously, including the spectral features of the background fit as priors. Including the background model uncertainties in the joint fit is important because subtracting the noisy background spectrum (Figure \ref{fig:flight_path_and_spectra}) would introduce considerable bias. Hence, the complete background model $b(E)$ consists of three Gaussian lines whose parameters are allowed to vary within the uncertainties of the background fit and a continuum whose amplitude and slope are left free. The model $b(E)$ is then multiplied by a global scaling factor $\beta$. The total sky model is the smooth simulated response to the DIRBE map multiplied by the unknown scaling parameter $\alpha$. Thus, this routine finds the best fit to the background and calculates the amplitudes $\alpha$ of the sky model and $\beta$ of the background model which best describe the signal region data. The final fit values of the continuum are $C_0$ = (11.3 $\pm$ 0.2) $\times$ 10$^{-4}$ cnts s$^{-1}$ keV$^{-1}$ and $\gamma$ = -4.1 $\pm$ 0.6. 

The fitted sky amplitude $\alpha$ = 1.1$^{+0.3}_{-0.3}$ and background amplitude $\beta$ = 28.0$^{+0.6}_{-0.7}$, reflecting the expected dominance of background. The signal-to-noise ratio is approximately 3.6. An $\alpha$ value of zero would indicate that the background model entirely explains the signal region data with no contribution from the sky model. A maximum likelihood ratio calculation yields a 3.7$\sigma$ significance above background. The line is found at an energy of $E_{\text{sky}}$ = 1811.3 $\pm$ 1.9 keV at a rate of 6.8 $\times$ 10$^{-4}$ cnts s$^{-1}$. Given the exposure time, this rate corresponds to an expected number of $\mathrm{^{26}Al}$ photons of $\sim$106. The background rate between 1803 and 1817 keV is 2.6 $\times$ 10$^{-3}$ cnts s$^{-1}$, implying $\sim$407 photons. The measured flux is discussed in Section \ref{sec:discussion_flux}. 

\begin{figure}
\centering
  \includegraphics[width=7cm,height=5cm]{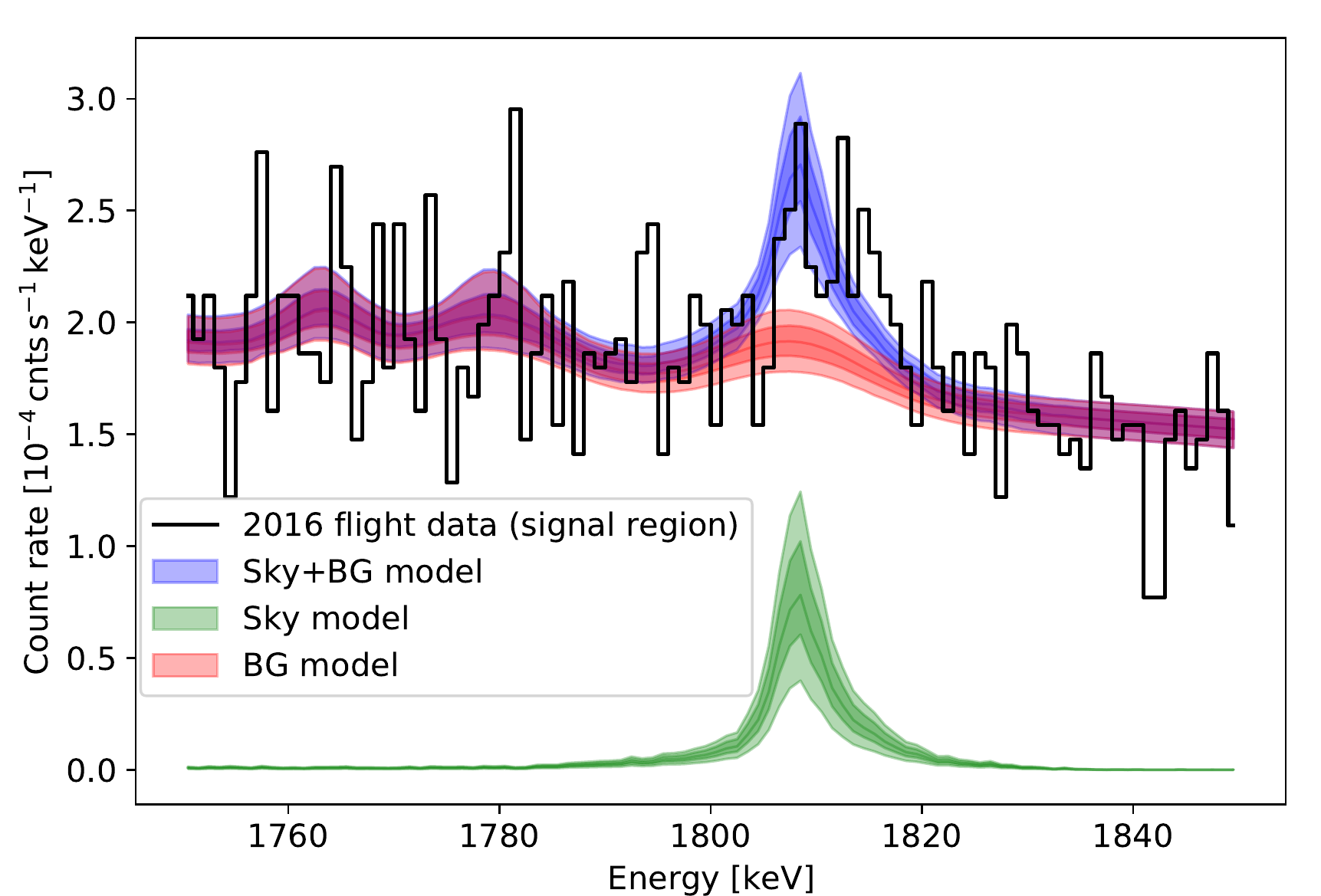}
  \includegraphics[width=8cm,height=5.45cm]{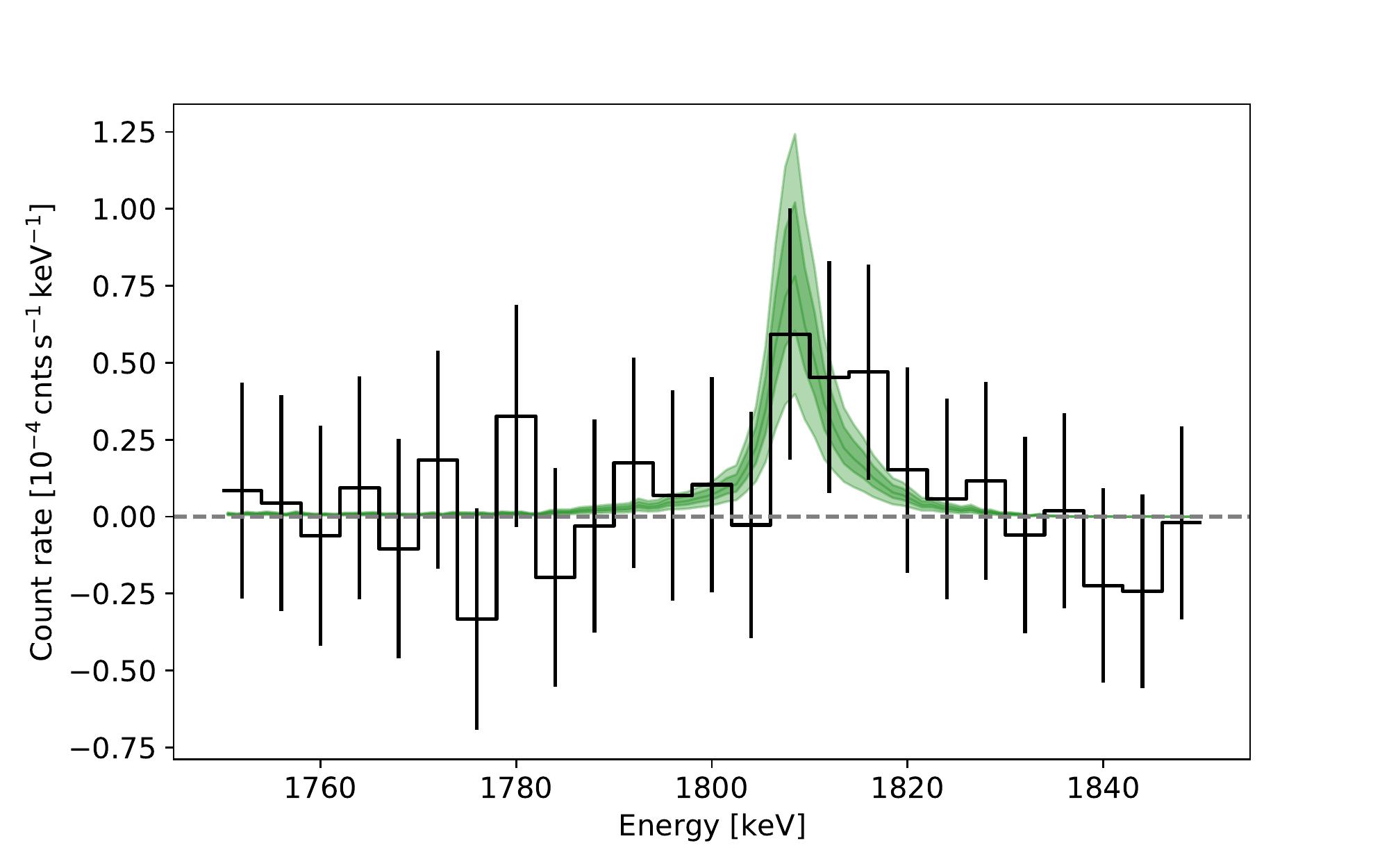}
\caption{Left: The total, sky, and background models plotted over the signal region spectrum. Right: The background-subtracted spectrum in 4-keV bins. The error bars are given by the expected variance of the fitted median sky model.}
\label{fig:flight_data_total_models_and_bg_sub_spec}
\end{figure}

Figure \ref{fig:flight_data_total_models_and_bg_sub_spec} shows the fitted models plotted over the signal region data. The central lines are the medians of the total, sky, and background models and the contours are 1$\sigma$ and 2$\sigma$ uncertainties. The contribution of the sky model to the peak in the signal region data near the desired $\mathrm{^{26}Al}$ 1.809 MeV line is visible. The background-subtracted spectrum in 4-keV bins is also shown in Figure \ref{fig:flight_data_total_models_and_bg_sub_spec}.

\section{Discussion} 
\label{sec:discussion_flux}
Taking into account the broadening of our signal region, the simulated response suggests a region larger than the Inner Galaxy (|$\ell|$ $\leq$ 30$^{\circ}$, |$b|$ $\leq$ 10$^{\circ}$). The measured flux from the COSI 2016 flight contained within this extended region is estimated to be (17.0 $\pm$ 4.9) $\times$ 10$^{-4}$ ph cm$^{-2}$ s$^{-1}$.

\begin{figure}
\centering
  \includegraphics[width=0.48\linewidth]{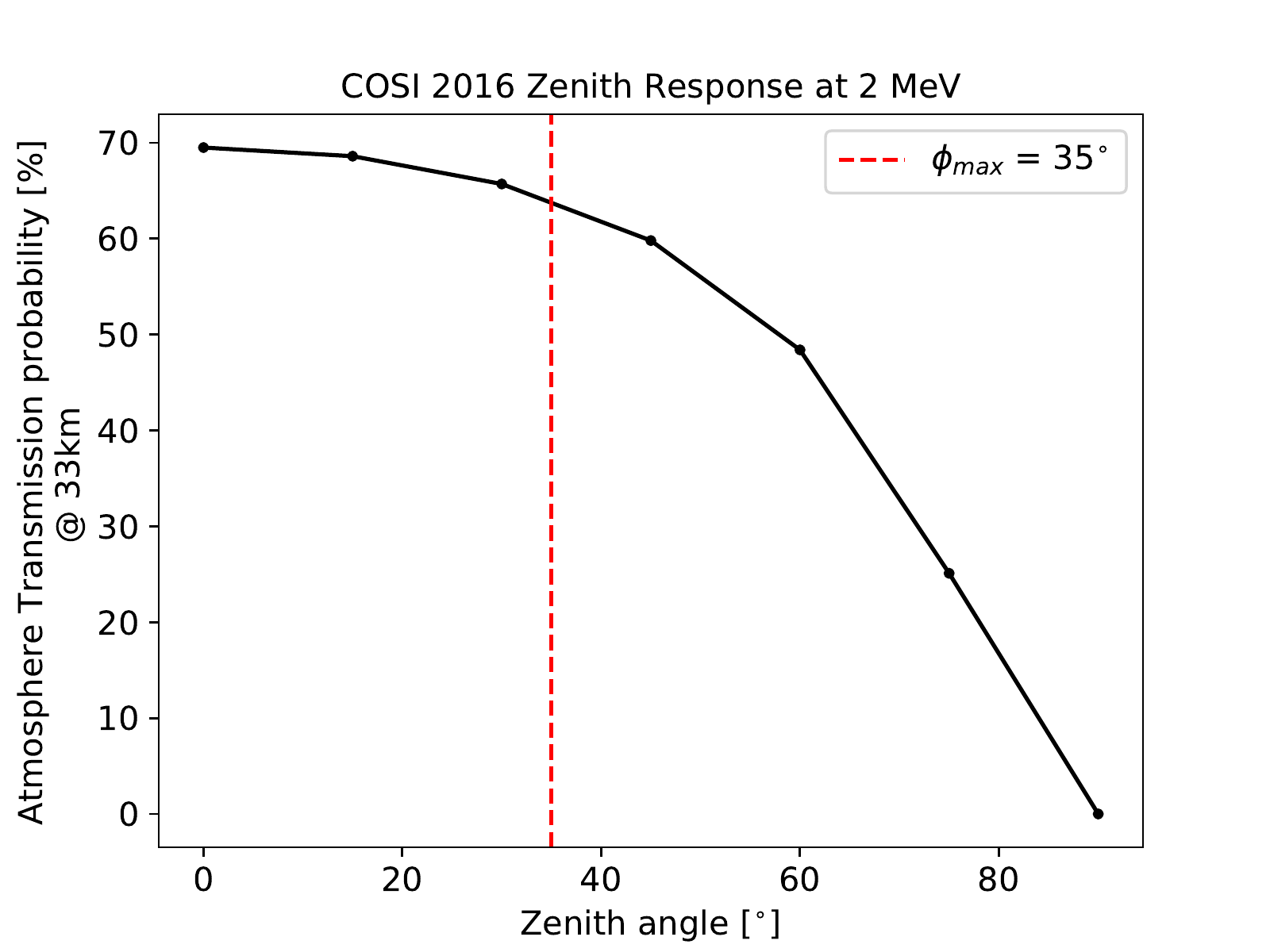}
  \includegraphics[width=0.48\linewidth]{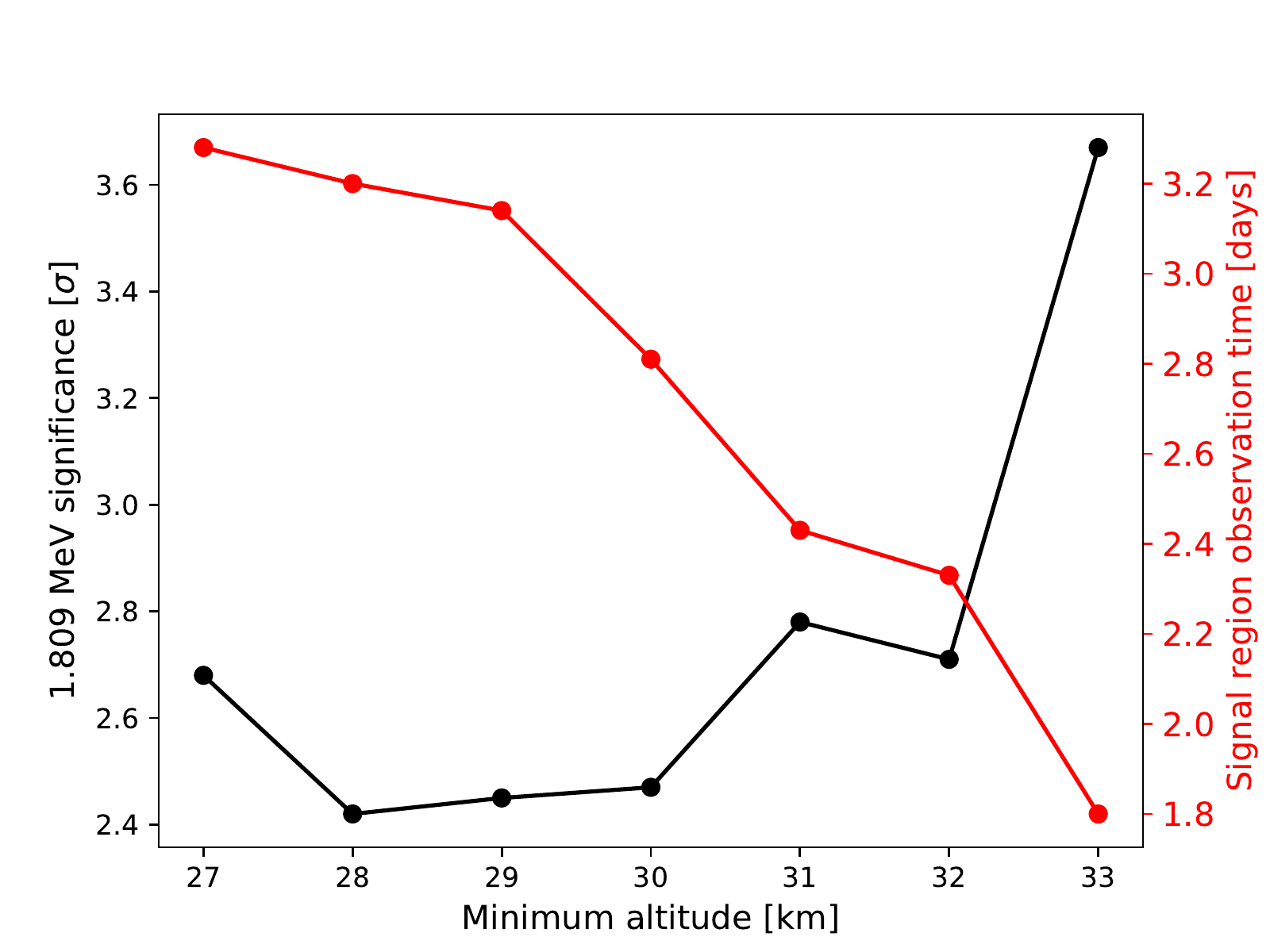}
\caption{Left: Zenith response of COSI to 2 MeV photons at the signal region flight altitude of 33 km. Right: Significance above background of the $\mathrm{^{26}Al}$ measurement  as a function of minimum flight altitude.}
\label{fig:zenith_resp_sig_vs_altitude}
\end{figure}

The extension beyond the Inner Galaxy is 35$^{\circ}$, the maximum Compton scatter angle employed in this analysis. Furthermore, the zenith response of COSI to 2 MeV photons at 33 km altitude indicates that the number of photons beyond an extension of 35--40$^{\circ}$ decreases exponentially (Figure \ref{fig:zenith_resp_sig_vs_altitude}). An expected flux value from the flight of 6.5 $\times$ 10$^{-4}$ ph cm$^{-2}$ s$^{-1}$ was calculated by integrating the flux, weighted by the zenith response, enclosed in the DIRBE 240 $\mu$m map over the 35$^{\circ}$-broadened region (|$\ell|$ $\leq$ 65$^{\circ}$, |$b|$ $\leq$ 45$^{\circ}$). The DIRBE map was normalized to contain the COMPTEL  Inner Galaxy $\mathrm{^{26}Al}$ flux of 3.3 $\times$ 10$^{-4}$ ph cm$^{-2}$ s$^{-1}$.

The SPI 1809 keV, COMPTEL 1809 keV, and ROSAT 0.25 keV \citep{rosat} maps were then examined for contributions to the signal data from extraneous photons. The ROSAT map is normalized to the full-sky flux of the SPI map, 1.7 $\times$ 10$^{-3}$ ph cm$^{-2}$ s$^{-1}$, and, contrary to the $\mathrm{^{26}Al}$ maps which exhibit strong emission in the Inner Galaxy, features emission concentrated towards high latitudes. It is thus a cross-check of our assumption of concentrated $\mathrm{^{26}Al}$ emission in the Inner Galaxy.

In the extended region, the weighted and integrated fluxes from SPI, COMPTEL, and ROSAT are 7.3 $\times$ 10$^{-4}$ ph cm$^{-2}$ s$^{-1}$, 8.2 $\times$ 10$^{-4}$ ph cm$^{-2}$ s$^{-1}$, and 3.2 $\times$ 10$^{-4}$ ph cm$^{-2}$ s$^{-1}$, respectively. The predominately high-latitude emission in ROSAT results in $\sim$50\% smaller flux than DIRBE.

Within $\sim$2$\sigma$ uncertainty, the COSI 2016 measured $\mathrm{^{26}Al}$ flux value is consistent with expectations from SPI and COMPTEL. It is approximately two times greater than that expected from simulations; this efficiency factor of two was also seen in COSI 2016 measurements of the 511 keV positron annihilation flux \cite{siegert2020imaging,kierans2020detection}. Applied to the 1.8 MeV line, this efficiency factor gives a flux of 8.5 $\times$ 10$^{-4}$ ph cm$^{-2}$ s$^{-1}$, fully consistent with expectations. Comparison with SPI and COMPTEL observations suggests that our systematic uncertainty on the overall flux normalization might be as high as 50\%. The consistency of our maximum-likelihood method is illustrated in Figure \ref{fig:zenith_resp_sig_vs_altitude}, which shows decreasing measurement significance with minimum flight altitude (increasing background). We conclude that we have measured a signal from the Inner Galactic region consistent with $\mathrm{^{26}Al}$. Investigations of the flight background with simulations are on-going to validate the results.

\bibliographystyle{JHEP}

\providecommand{\href}[2]{#2}\begingroup\raggedright\endgroup

\clearpage
\section*{Full Authors List: \Coll\ Collaboration}

\scriptsize
\noindent
Jacqueline Beechert$^1$, 
Thomas Siegert$^2$, 
Andreas Zoglauer$^1$, 
Alexander Lowell$^1$,
Carolyn Kierans$^3$,
Clio Sleator$^4$,
Hadar Lazar$^1$,
Hannah Gulick$^1$,
Jarred M. Roberts$^{5}$,
John Tomsick$^1$,
Peter von Ballmoos$^6$,
Pierre Jean$^6$,
Steven E. Boggs$^{5,1}$,
Theresa J. Brandt$^3$ \\

\noindent
$^1$Space Sciences Laboratory, University of California, Berkeley, 7 Gauss Way, Berkeley, CA 94720, USA.
$^2$Institut f{\"u}r Theoretische Physik und Astrophysik, Universit{\"a}t W{\"u}rzburg, Campus Hubland Nord, Emil-Fischer-Str. 31, 97074 W{\"u}rzburg, Germany.
$^{3}$NASA Goddard Space Flight Center, Greenbelt, MD 20771, USA.
$^{4}$U.S. Naval Research Laboratory, Washington, DC 20375, USA.
$^{5}$Center for Astrophysics and Space Sciences, UC San Diego, 9500 Gilman Drive, La Jolla CA 92093, USA.
$^{6}$IRAP, 9 Av colonel Roche, BP44346, 31028 Toulouse Cedex 4, France.

\end{document}